\documentclass{article}

\usepackage{arxiv}

\usepackage[utf8]{inputenc} 
\usepackage[T1]{fontenc}    
\usepackage{hyperref}       
\usepackage{url}            
\usepackage{booktabs}       
\usepackage{amsfonts}       
\usepackage{nicefrac}       
\usepackage{microtype}      
\usepackage{lipsum}
\usepackage{graphicx} 
\usepackage{float,subfigure}

\title{Viscosity of macromolecules with complex architecture}

\author{
  Vladimir Yu. Rydyak \\
  Faculty of Physics\\
  Lomonosov Moscow State University\\ Moscow, 119991, Russia\\
  \texttt{vurdizm@gmail.com} \\
   \And
 Artem V. Sergeev \\
  Semenov Federal Research Center for Chemical Physics\\
  Moscow, 119991, Russia \\
  \And
 Elena Yu. Kozhunova \\
  Faculty of Physics\\
  Lomonosov Moscow State University\\ Moscow, 119991, Russia\\
  \And
 Alexander V. Chertovich \\
  Semenov Federal Research Center for Chemical Physics\\
  Moscow, 119991, Russia\\
  Faculty of Physics\\
  Lomonosov Moscow State University\\ Moscow, 119991, Russia\\
}

\begin{document}
\maketitle

\begin{abstract}
It is well-known that the architecture of macromolecules plays an important role in the hydrodynamics and viscosity of its semi-diluted solutions. However, the systematic study of the rheology of macromolecules with complex topology, such as microgels, remains a difficult task. In this work, we use the computer simulations methods of non-equilibrium molecular dynamics to study the viscous properties of randomly cross-linked microgels and compare it to multiple reference systems (linear chains, 10-arm star polymers and hard spheres). We show that the microgel cross-linking density and, thus, the particle shape, plays crucial role in its viscosity. Also, in contrast to a simpler polymer systems, microgel dispersions in good solvent show much less shear thinning.
\end{abstract}


\section{Introduction}
The topic of macromolecule suspension of various architecture is of high importance both from fundamental and practical points of view and therefore was discussed in literature numerous times. That interest is backed by a very wide range of application of polymer solutions: from painting materials to medical use. Recently the controlled polymerization techniques made it possible to accurately manage the molecular weight and architecture of macromolecular objects such as stars, dendrimers, combs etc. All these classes have different dependencies of rheological properties on the molecular weight (solution viscosity, gyration radius and aspect ratio). The two extreme cases, linear chains and hard spheres, are well studied both experimentally and theoretically.      

The first analytical derivation for viscosity of spherical particles solutions were presented by Einstein \cite{Einstein1906}: $\eta = \eta_0 (1+2.5\phi)$, where $\eta_0$ is solvent viscosity and $\phi$ is solute volume fraction. This linear approximation provides acceptable accuracy only for low concentration. For more concentrated solution (up to $\phi=0.2$) there is a second order approximation \cite{batchelor1977effect}: $\eta = \eta_0 (1+2.5\phi+6.2\phi^2)$. A general expression for solution of incompressible particles of arbitrary shape was also proposed \cite{krieger1959mechanism}:  $\eta = \eta_0 (1-\phi/\phi_{max})^{-[\eta]\phi_{max}}$, where $\phi_{max}$ is dense packing volume fraction ($\phi_{max}=0.74$ for ideal rigid spheres), $[\eta] = \lim_{c\rightarrow 0} {(\eta-\eta_0)/\eta_0c}$ is so-called intrinsic viscosity, and $c$ is solute concentration. For example, elipsoids of aspect ratio $\delta$ show the following phenomenological dependence \cite{brenner1974rheology, pabst2006particle}: $[\eta]=2.5+0.123(\delta -1)^{0.925}$. The Mark–Houwink equation is widely used to estimate $[\eta]$ in polymer solutions \cite{rubinstein2003polymer}: $[\eta]=KM^{\alpha}$, where $M$ is polymer molecular weight, $K$ and $\alpha$ are constants for a specific polymer-solvent pair. More sophisticated microscopic models also help to predict solution viscosity. First of all, we should note Kirkwood theory \cite{kirkwood1948intrinsic} and Rouse model \cite{rouse1953theory}, later modified by Zimm \cite{zimm1956dynamics}. However, Zimm model does not predict the viscosity dependence on the shear rate, which contradicts experimental data. Moreover, the generalization of Zimm model for non-linear polymers (stars, combs etc.) is excessively complicated and adopts empirical assumptions \cite{lu2013intrinsic}. 
Finally, the above mentioned models are applicable only for low concentrations (as even the intrinsic viscosity itself is defined for the zero-limit concentration) and hardly can be used to predict viscosity of macromolecule solutions on practice.     
Numerical simulations and particularly molecular dynamics (MD) provide a robust tool to investigate polymer solution rheological properties at molecular scale. It is particularly useful at conditions out of reach for theory: complex molecular architectures and interactions or high concentrations. An excellent example is thorough study of linear polymers solution rheology by Aust et al. \cite{Aust1999}. Special attention was given to the dependence of the solution properties on the shear rate and shear thinning effect. In \cite{khabaz2014effect,yao2020combined}, computer simulation was employed to compare various macromolecule architectures including rings, stars, combs and H-like polymers. The modeled solution concentration was limited by overlap concentration which made it possible to relate the simulation results with existing theoretical approximations.   

 Microgels are the example of complex polymer objects with random topology, much more sophisticated than stars and simple cycles. Microgels, especially environment-sensitive, are widely studied by the scientific community for the last decade and start to find its place in various applications and technologies. The main features of this species are the following: characteristic size of microgels is from tens to hundreds of nanometers; tunable radial density distribution, such as loosely cross-linked particles, fuzzy spheres, or compact dense networks; the ability to drastically change the overall particle volume and topology in response to external stimuli, especially temperature. It seems impossible to construct a theoretical model of rheological properties of such complex objects, and computer simulations are required. Although molecular simulation tools were applied to microgels objects before (e.g. \cite{martin2019review}), its behaviour in shear flow condition was not considered yet.

\begin{figure}[H]
\centering
\includegraphics[width=0.7\linewidth]{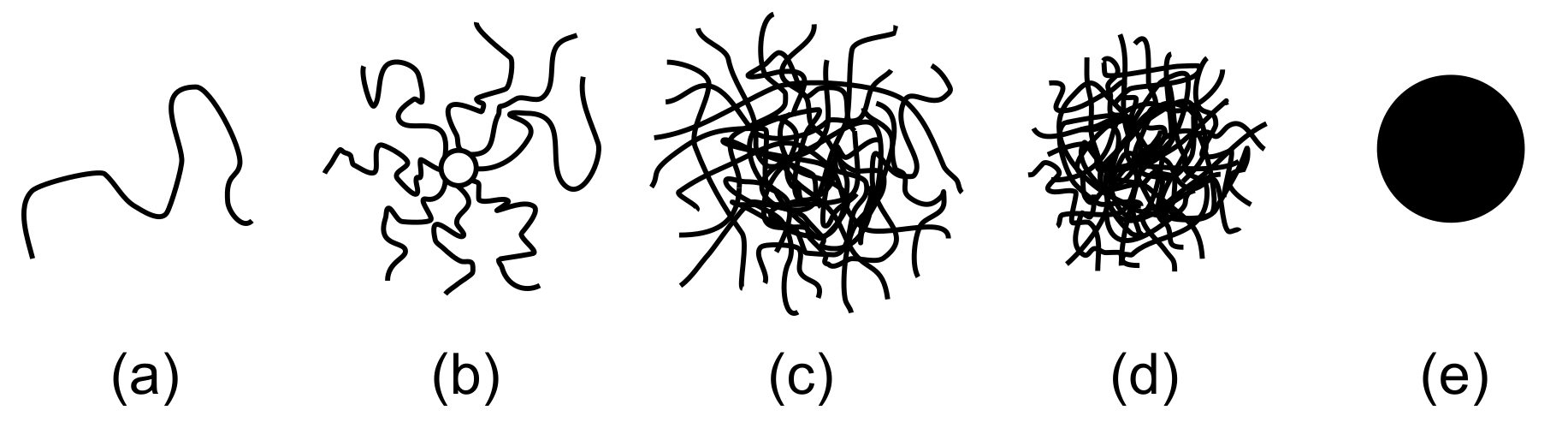}
\caption{Schematic representation of the systems studied in this work: (a) linear polymers, (b) star-like polymers, (c) low cross-linked microgels, (d) highly cross-linked microgels, and (e) hard spheres}
\label{fig:systems}
\end{figure}

In this work we employ non-equilibrium molecular dynamics methods to study behaviour of microgel solution at relatively high concentration 16\% and to compare it with other macromolecule classes: linear chains, stars and dense spheres (see Fig.\,\ref{fig:systems}). A special attention is given to the dependency of rheological properties such as viscosity, gyration radius and aspect ratio on the molecular wight. We show that microgel solutions exhibit both features typical for hard spheres and polymers. Moreover its behaviour can be tuned by controlling the size and cross-linker content, which makes microgels a promising flexible object for material design.

\section{Methodology}

Non-equilibrium molecular dynamics (NEMD) simulation approach was used in this work to investigate rheological properties of polymer solutions. The polymer chains were represented by bead-spring model employing the finitely extensible nonlinear elastic (FENE) potential \cite{kremer1990dynamics} for the interaction of connected beads. The potential is given as:
$$E = -0.5KR_0^2ln[1-(r/R_0)^2],$$
where the parameters were set as $R_0 = 1.5$ and $K = 30.0$.  The solvent was modeled explicitly by free-floating individual beads equivalent to the polymer beads. All particles was subjected to 12--6 Lennard-Jones potential with both $\epsilon$ and $\sigma$ parameters set to 1.0, and cutoff distance $r_{cut} = 2^{1/6}\sigma$ (purely repulsive interaction). Such interaction model corresponds to a polymer solution in a ``good'' solvent. All molecular dynamics simulations were performed using the LAMMPS software package \cite{plimpton1995fast}. The temperature was kept constant at $T = 1.0$ by employing the Nosé–Hoover thermostat \cite{shinoda2004rapid}. The integration time-step of all productive runs was 0.005. In all cases the total density of the system was 1.0, the simulation box dimensions were $100\times100\times100$ (i.e. the box contained 1 million particles). The polymer solution viscosity values were calculated using NEMD simulations with Lees-Edwards boundary  conditions \cite{Lees1972, j2007statistical, tuckerman1997modified} that implies simulation box deformation in order to simulate gradient flow velocity. In most of the simulations (until otherwise mentioned) the constant engineering shear strain rate was set to $\gamma=0.001$ which results in the total flow velocity difference across the cell equal to 0.1 (as box size is 100).

In this study linear polymers were represented by chains consisting of 50, 75, 100 and 150 beads. The 10-armed star polymers with branch lengths 50, 100 and 200 beads were constructed by attaching linear chains to a central ring containing 10 beads rather than by attaching all the arms to a single central bead.

The microgels were prepared via simulations of precipitation polymerization \cite{Rudyak2019}. In this work, we used  microgels with 1\%, 5\% and 10\% cross-linker (CL) content. The higher the cross-linker content, the denser and less deformable the microgel particle, which makes it more similar to hard spheres. To study concentration dependency of viscosity, we used microgels with 5\% and 10\% CL with $M=1000$. To study the impact of CL content on viscosity, we used 1\% CL microgels of molecular mass $M$ equal to 5000, 10000 and 20000 (lower $M$ leaded to poorly cross-linked non-representative structure) and 5\% CL microgels with $M$ of 5000 and 10000. These systems were studied only at concentration $c=0.16$, as lower concentrations would have contain too small number of microgel particles to study rheological behavior. 

Hard spheres of molecular mass 100, 1000 and 10000 were prepared by cutting spherical part from a diamond-like lattice and forming additional bonds between neighbour particles in it to improve its non-deformability. 

For all objects, in order to prepare the initial system, the polymer molecules were placed in a large simulation cell without solvent. Then the cell was slowly compressed towards the box size $100\times100\times100$ and the polymer molecules were allowed to relax long enough for the average values of gyration radius components to reach the stationary values (several million MD steps). Then, a home-made algorithm was used to insert solvent particles into the void spaces to reach the desired total density (1.0), and finally the system was equilibrated for 1 million MD steps.

\section{Results and Discussions}

\subsection{Effect of shear rate}

Polymer solutions belong to the class of shear-thinning fluids as its viscosity decreases with increase of shear rate above some critical shear rate value \cite{bird1987dynamics}. The dependence of viscosity $\eta$ on the shear rate $\gamma$ is well described by expression $\eta(k) = \eta_0\exp(-\alpha \gamma)$ \cite{Phillies1999} which is true for at least linear and star polymers. The critical shear rate value is defined by polymer relaxation time thus depends on both molecular weight and topology. Although it seems most straightforward to compare viscosity values at low shear rate limit, there are serious technical issues. In simulations the viscosity value is calculated based on averaged momentum flux in the direction of flow velocity gradient. At low imposed shear rate the relative amplitude of momentum flux fluctuations becomes larger (i.e. signal to noise ratio decreases) so that the simulation time must be increased in order to obtain enough data points. Thus in case of systems with large relaxation times the below critical shear rate condition would demand an unreasonable amount computational resources.  

\begin{figure}[htb]
\centering
\subfigure{
  \includegraphics[width=0.45\linewidth]{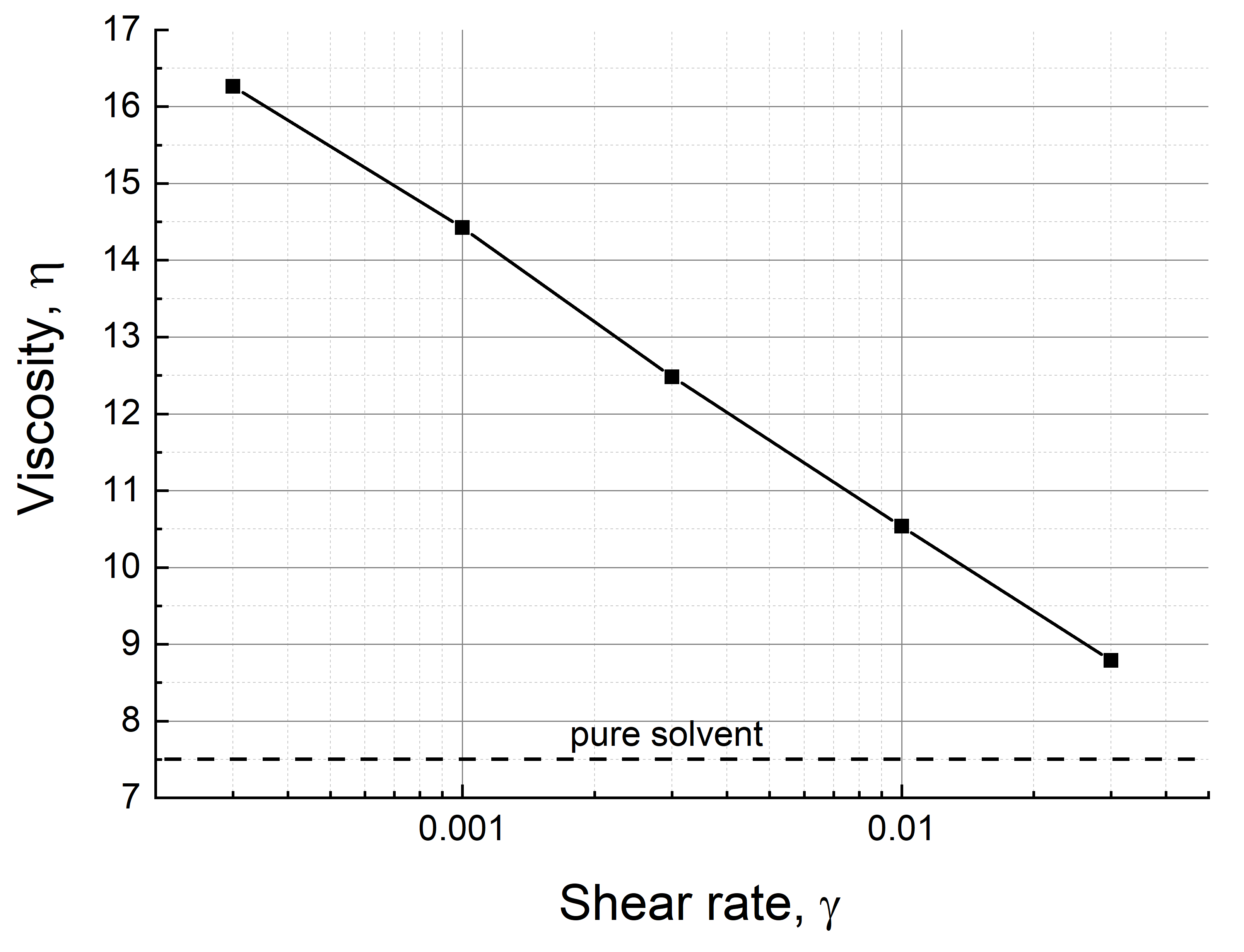}}
\subfigure{
  \includegraphics[width=0.45\linewidth]{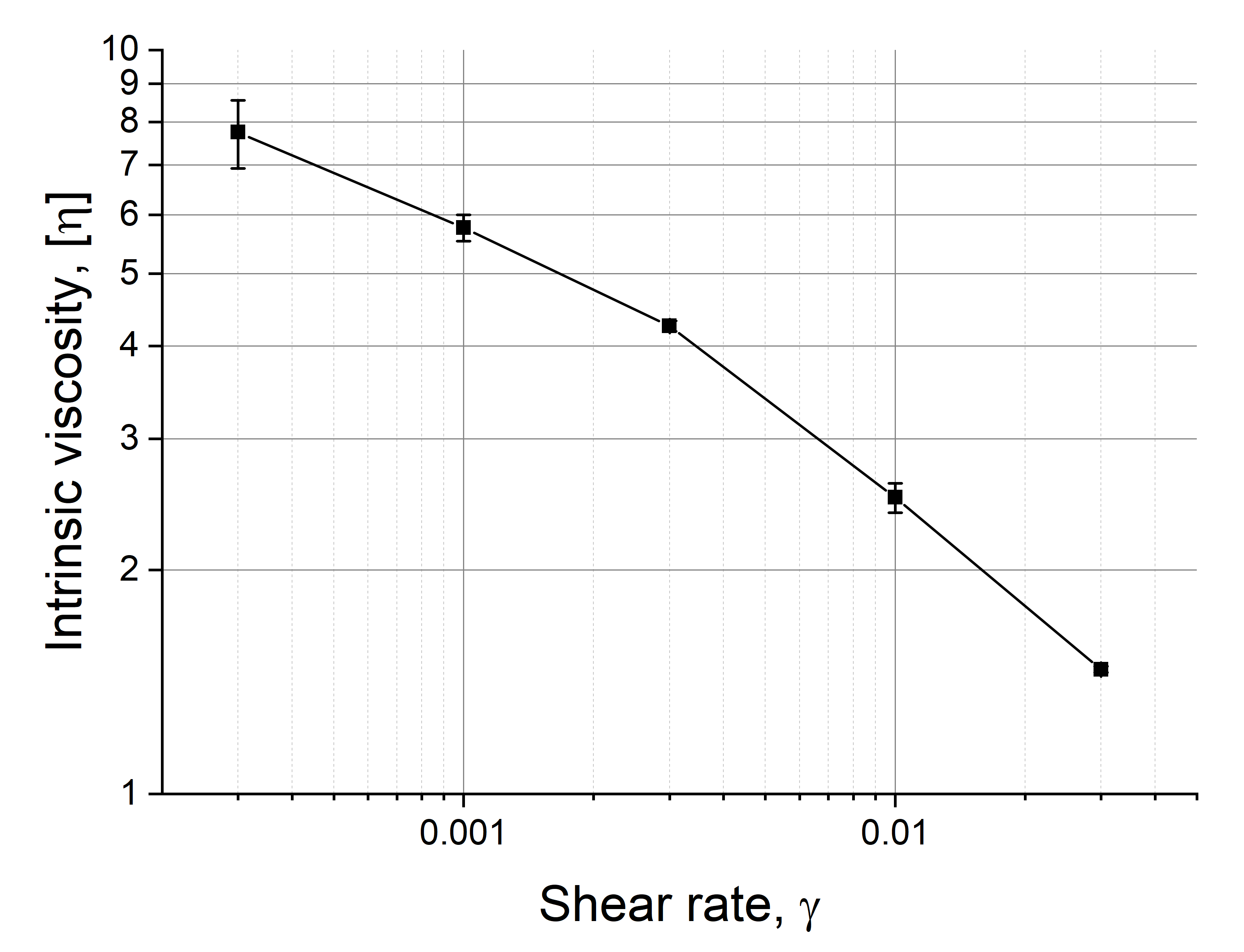}}
\caption{Absolute viscosity $\eta$ (a) and intrinsic viscosity $[\eta]$ (b) as functions of shear rate for the solution of $M = 50$ linear chains at 16\% concentration.}
\label{fig:gamma-visc}
\end{figure}

In order to choose the shear rate value we have calculated the viscosity of the solution of linear chains of length $M=50$ at concentration $c = 16\%$ and various shear rates $\gamma$. Fig.\,\ref{fig:gamma-visc} shows the dependencies of intrinsic viscosity $[\eta]$ on shear rate $\gamma$. These results are in agreement with the previously known data \cite{Kotaka1966,Phillies1999}, which indicates the correct behaviour of our methodological setup. For the following simulations, we have fixed the value of the shear rate at $\gamma=0.001$ as a reasonable compromise between shear thinning regime and requirements of accurate simulations in a reasonable time on modern supercomputers.

\subsection{Viscosity of microgels}

The inclusion of microgel molecules into solution causes viscosity growth for the two reasons: (1) microgels act as large semi-hard particles, and (2) at high enough concentrations, the entanglements between dangling ends in the coronas of neighbour microgels are formed. The first is defined mostly by the overall size of the molecule (expressed in radius of gyration $R_g$). The second is strongly related to the amount of cross-linker in the microgel, as it directly affects the average length of the dangling ends in the microgel and its the corona width \cite{Rudyak2019}.

\begin{figure}[htb]
\centering
\subfigure{
  \includegraphics[width=0.45\linewidth]{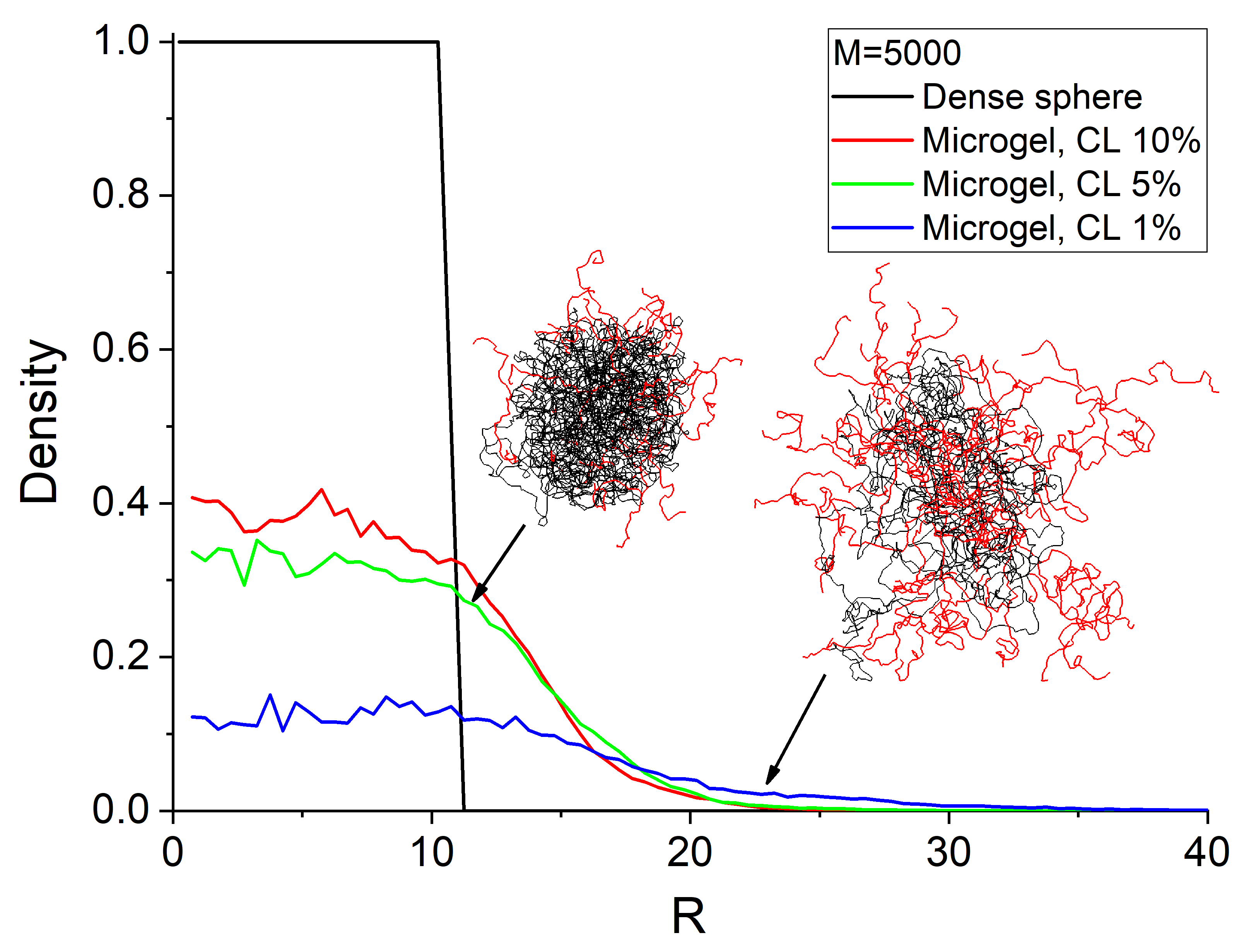}
  \label{fig:mg-char}}
\subfigure{
  \includegraphics[width=0.45\linewidth]{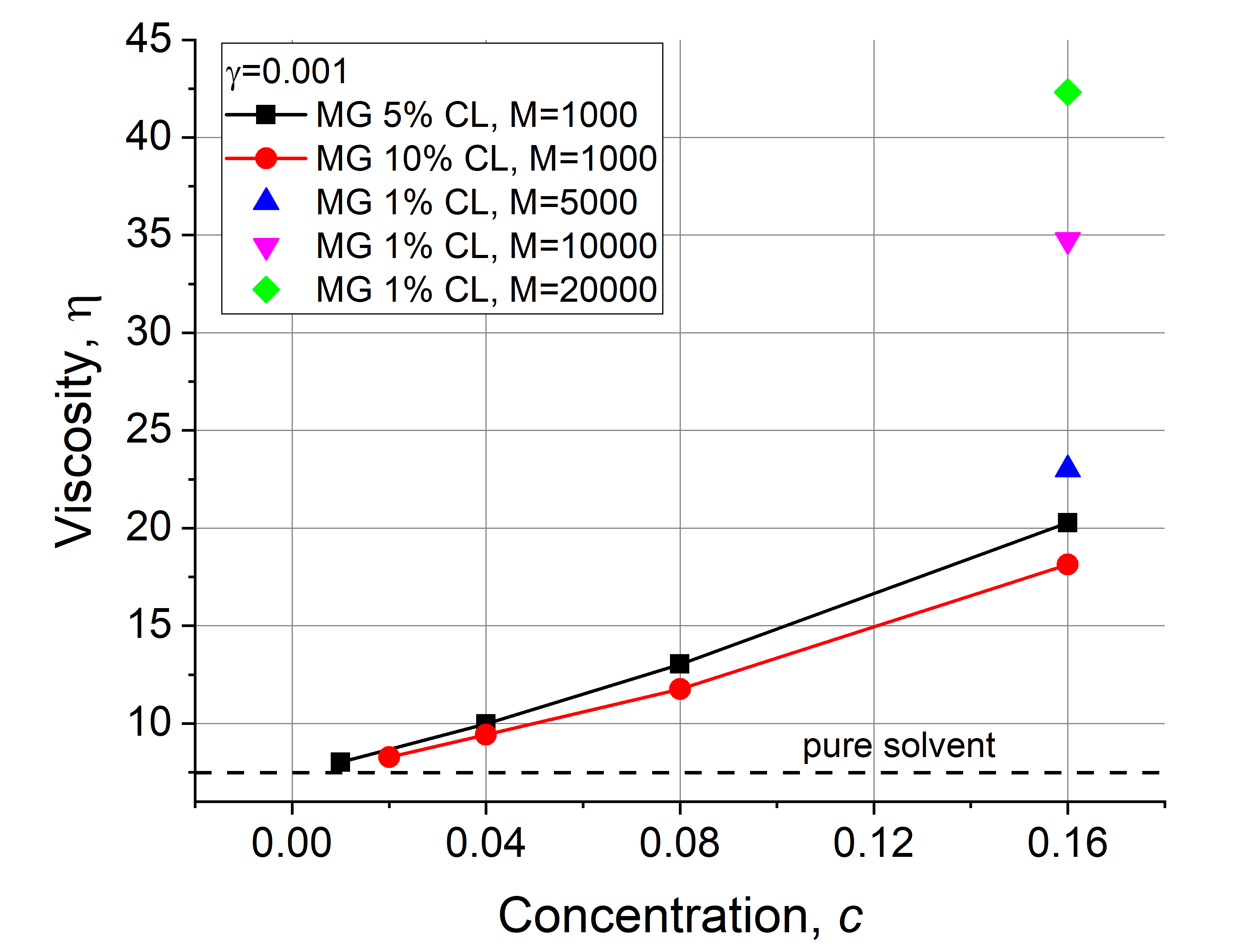}
  \label{fig:mg-visc}}
\caption{(a) Viscosity of microgel solutions with various concentrations and  cross-linker content in microgels. (b) Density profile of microgels with various cross-linker content and hard sphere ($M=5000$). 1\%\,CL  and 5\%\,CL microgel images on inset are colored in black (network part) and red (dangling ends).}
\label{fig:mg-data}
\end{figure}

Fig.\,\ref{fig:mg-char} shows radial density distribution for microgels with $M=5000$ and CL content from 1\% to 10\% as well as for hard sphere of equal mass. The tail in the density profile depicting corona is significantly larger for 1\% CL microgel in comparison with 5\% and 10\% CL particles. On the insets, dangling ends forming the corona are marked in red for both particles, while the network part is marked with black. For 5\% and 10\% CL microgels, the amount and length of dangling ends are too low to play any significant role on rheological properties. At the same time, 1\% CL microgels exhibit large amount of dangling ends, which average length is higher for more massive particles. 

Consequently, the rheological behaviour of microgels with high and low CL content are different. Fig.\,\ref{fig:mg-visc} shows the dependency of the system viscosity on both microgel concentration and the amount of cross-linker in the microgel (for 5\% and 10\% CL microgels with $M=1000$). The small microgels with high cross-linker content affect on the total viscosity almost linearly with concentration up to as high as $c=0.16$. This is because the dangling ends are too short and too few in these microgels, and the entanglements are easy to break during the shear flow. At the same time, the microgels with 1\% CL demonstrate much higher viscosity, and the resulting viscosity grows significantly with the increase of the mass of microgel molecule.

\subsection{Influence of molecular mass on viscosity}

\begin{figure}[htb]
\centering
\subfigure{
  \includegraphics[width=0.46\linewidth]{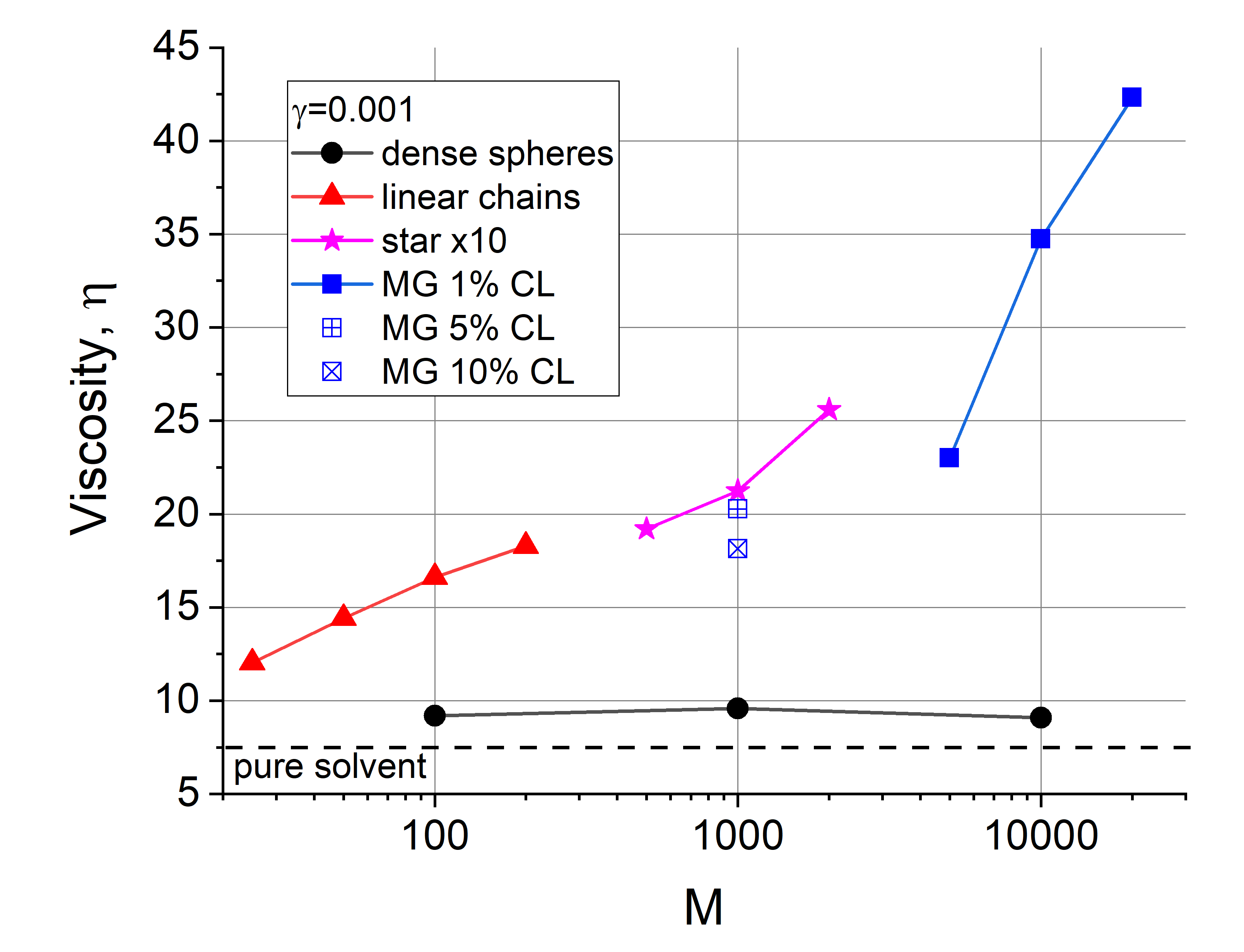}
  \label{fig:mass-visc}} 
\subfigure{
  \includegraphics[width=0.45\linewidth]{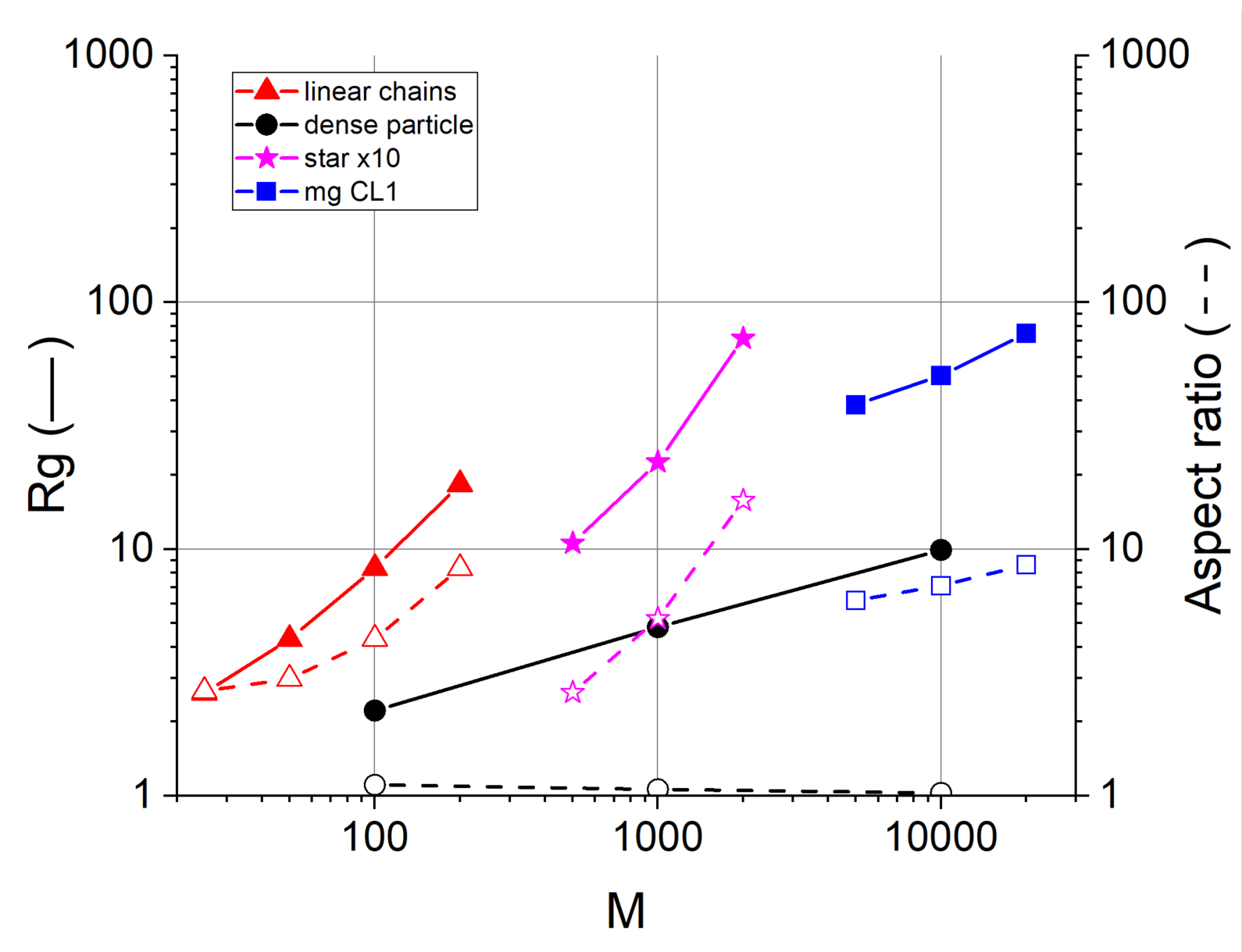}
  \label{fig:mass-char}}
\caption{Absolute viscosity as a function of molecular mass $M$ (a), radius of gyration ($R_g$) and aspect ratio as functions of molecular mass $M$ (b). Solute concentration $c=0.16$.}
\label{fig:mass-data}
\end{figure}

We have calculated solution viscosity for the three other systems: linear polymers (chain length from 25 to 200), 10-branched star polymers (branch length from 50 to 200), and hard spheres (from 100 to 10000 particles). All systems were studied at concentration $c=0.16$, the resulting viscosity data are shown in Fig.\,\ref{fig:mass-visc}. 

The size of the hard spheres does not affect the viscosity in the studied range. That is in agreement with the theoretical derivations for spherical particles \cite{batchelor1977effect, krieger1959mechanism}. This case can be considered as a reference for the lowest solution viscosity at fixed polymer concentration. 
 
The viscosity of linear polymers  appears to slowly increase with chain length. Relying on the Mark-Houwink equation, one could expect a steeper increase rate. However, shear rate $\gamma = 0.001$ gives rise to the shear thinning effect that mitigates the influence of the molecular weight \cite{Aust1999}. The longer the chain the higher its tendency to orient and elongate along the flow direction, which decreases the inter-chain entanglement probability and impedes further viscosity growth. It can be expressed in terms of the average aspect ratio calculated as the ratio between maximum and minimum terms of a diagonalized tensor of inertia.High aspect ratio values can bee seen in Fig. \ref{fig:mass-char}, and the scaling exponent of the gyration radius about 0.94 is close to linear object (the theoretical value for Gaussian chains equals to 0.588 \cite{flory1953principles}).  

The above mentioned is also true for star polymers, bearing in mind that the molecular weight range is by an order of magnitude higher due to multiple (ten) arms, while the linear size of stars (i.e. arm length) is of the same order as for the linear polymers in our calculations. The similarity of behaviour between the linear and star polymers was also revealed in \cite{khabaz2014effect, khasat1988dilute, picot2007solutions}.

In contrast, the viscosity of 1\% CL microgel solution grows significantly with particle mass. Moreover, it grows much faster than for any other objects covered in this study. We believe this can be explained by the following. From one side, microgels are dense enough to be difficult to deform by the shear flow. According to data presented in Fig.\ref{fig:mass-char}, the  radius of gyration ($R_g$) scaling exponent for 1\% CL microgels is significantly smaller than that for the linear or star polymers and very close to that of the hard spheres. The aspect ratio weakly depends on the molecular weight and lies below 10 for mass up to 20000. From another side, dangling ends forming corona of the microgel become longer upon increasing the molecular weight \cite{Rudyak2019}. It promotes inter-chain entanglement and causes raise in the viscosity. 

\begin{figure}[htb]
\centering
  \includegraphics[width=0.95\linewidth]{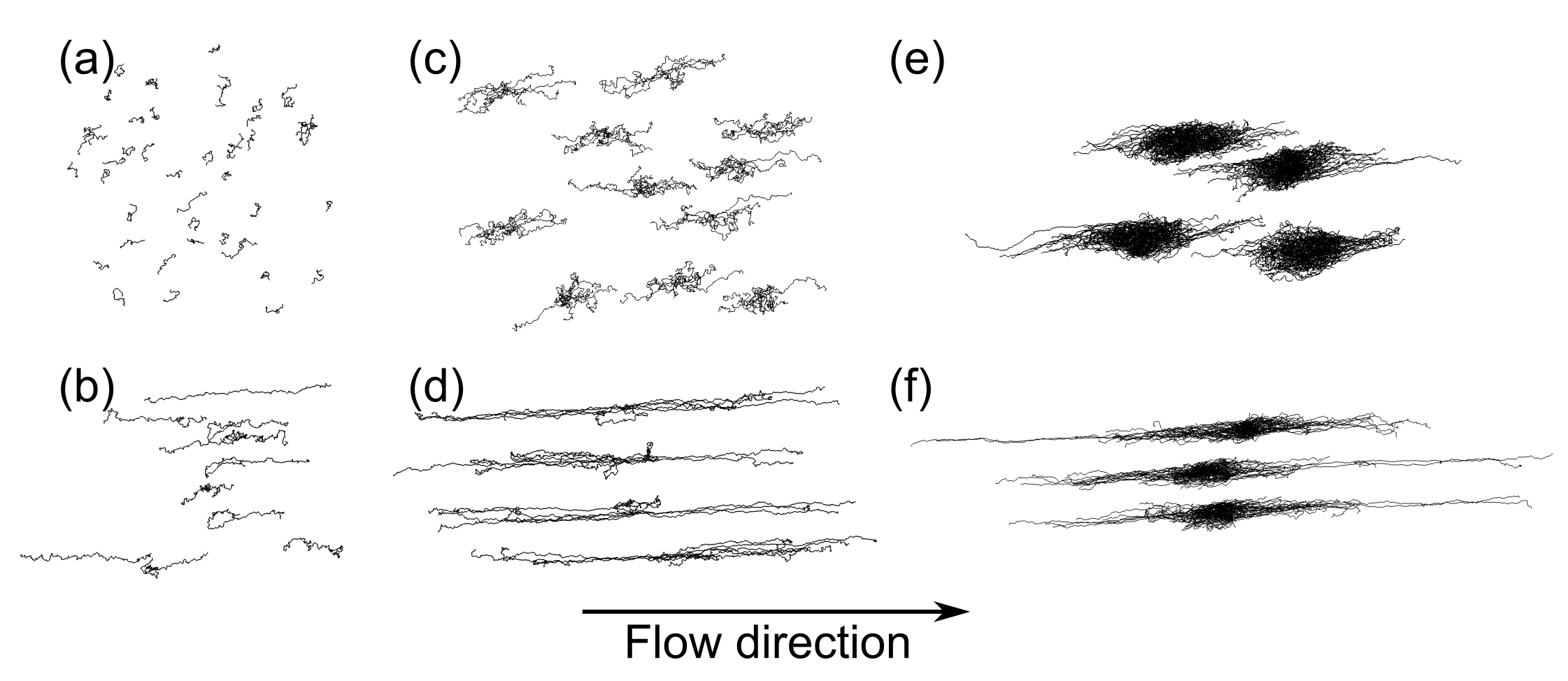}
  \label{fig:spanshots}
\caption{Snapshots of equilibrium conformations under shear flow $\gamma=0.001$ for linear polymers of chain lengths 25 (a) and 200 (b), star polymers of side chain length 50 (c) and 200 (d), 2\% CL microgels with $M=10000$ (e), and 1\% CL microgels with $M=10000$ (f).}
\label{fig:mass-data}
\end{figure}

\section{Conclusions}

In this work we investigated the dynamic viscosity of solutions of macromolecules with different architecture using computer simulations. We used non-equilibrium molecular dynamics with Lees-Edwards periodic boundaries on the open-source LAMMPS platform. For the first time, we evaluated in simulations the viscoelastic behaviour of microgels depending on its topology and size.  We studied the viscosity of swollen microgels with 1\%, 5\% and 10\% cross-linker content as well as multiple reference systems (linear chains, 10-arm star polymers and hard spheres). We found that the microgel cross-linker content plays crucial role in its viscosity due to the influence on formation of dense core and loose corona consisting of dangling ends. This allows to obtain low-viscosity solutions of polymer systems with high molecular mass. Also stretching of microgels along the flow direction is much less pronounced  because of cross-linked core. This feature allows microgels to have more stable dynamic viscosity and less pronounced shear-thinning effect. Since the topology of real microgel particles can be controlled on the synthesis stage, the viscosity properties of the resulting dispersion is adjustable according to the requirements of application. To summarize, the proposed methodology for modeling the rheology of microgels can help to design the particles with suitable viscosity properties.

Note that a study of macromolecular solution viscosity in simulations has become possible only recently, just ten years ago the available computing power did not allow us to do such research. Now this has become a reality and, we are sure, new results with macromolecules of a different architecture (ridges, ladder, dendritic, etc.), as well as work on larger microgels and even polymer micelles will soon spring up.

\section*{Acknowledgments}
This research was funded by Russian Foundation for Basic Research (Grant No. 19-33-70052).
The research was carried out using the equipment of the shared research facilities of HPC computing resources at Lomonosov Moscow State University.

\bibliographystyle{unsrt}  


\end{document}